\documentclass[prd,preprint,superscriptaddress,preprintnumbers,eqsecnum,showpacs,nofootinbib,nobibnotes]{revtex4-1}
\usepackage{amsfonts,amsmath,amssymb,bm}
\usepackage[table]{xcolor}
\usepackage{graphicx} 
\usepackage{mathtools}
\usepackage{boondox-cal}
\usepackage{xcolor}

\usepackage[utf8]{inputenc}
\usepackage{slashed}

\newcommand{\be}{\begin{equation}}
\newcommand{\bea}{\begin{eqnarray}}
\newcommand{\ba}{\begin{align}}
\newcommand{\ee}{\end{equation}}
\newcommand{\eea}{\end{eqnarray}}
\newcommand{\ea}{\end{align}}

\definecolor{zero2}{rgb}{0.88,0.88,.88}

\def\1eq#1{Eq.~(\ref{#1})}

\def\2eqs#1#2{Eqs.~(\ref{#1}) and~(\ref{#2})}
\def\3eqs#1#2#3{Eqs.~(\ref{#1}),~(\ref{#2}) and~(\ref{#3})}
\def\4eqs#1#2#3#4{Eqs.~(\ref{#1}),~(\ref{#2}),~(\ref{#3}) and~(\ref{#4})}

\def\tbarc{\bar{\mathcal{c}}^*}
\def\tT{\mathcal{T}_1}
\def\tU{\mathcal{U}}
\def\tR{\mathcal{R}}

\def\rc#1{\rho_{#1}}
\def\trc#1{{\tilde \rho}_{#1}}

\def\op#1{{\cal O}_{#1}}

\def\s#1{{\scriptscriptstyle #1}}

\def\G{\Gamma}

\def\T{T_1}

\def\s{\mathcal{s}}

\def\hphi0{{\hat\phi}_0}

\def\d{\!\mathrm{d}^4x\,}

\def\bkg#1{\widehat{#1}}

\makeatletter
\def\user@resume{resume}
\def\user@intermezzo{intermezzo}
\newcounter{previousequation}
\newcounter{lastsubequation}
\newcounter{savedparentequation}
\renewenvironment{subequations}[1][]{%
      \def\user@decides{#1}%
      \setcounter{previousequation}{\value{equation}}%
      \ifx\user@decides\user@resume 
           \setcounter{equation}{\value{savedparentequation}}%
      \else  
      \ifx\user@decides\user@intermezzo
           \refstepcounter{equation}%
      \else
           \setcounter{lastsubequation}{0}%
           \refstepcounter{equation}%
      \fi\fi
      \protected@edef\theHparentequation{%
          \@ifundefined {theHequation}\theequation \theHequation}%
      \protected@edef\theparentequation{\theequation}%
      \setcounter{parentequation}{\value{equation}}%
      \ifx\user@decides\user@resume 
           \setcounter{equation}{\value{lastsubequation}}%
         \else
           \setcounter{equation}{0}%
      \fi
      \def\theequation  {\theparentequation  \alph{equation}}%
      \def\theHequation {\theHparentequation \alph{equation}}%
      \ignorespaces
}{%
  \ifx\user@decides\user@resume
       \setcounter{lastsubequation}{\value{equation}}%
       \setcounter{equation}{\value{previousequation}}%
  \else
  \ifx\user@decides\user@intermezzo
       \setcounter{equation}{\value{parentequation}}%
  \else
       \setcounter{lastsubequation}{\value{equation}}%
       \setcounter{savedparentequation}{\value{parentequation}}%
       \setcounter{equation}{\value{parentequation}}%
  \fi\fi
  \ignorespacesafterend
}
\def\CT@@do@color{%
	\global\let\CT@do@color\relax
		\@tempdima\wd\z@
		\advance\@tempdima\@tempdimb
		\advance\@tempdima\@tempdimc
		\advance\@tempdimb\tabcolsep
		\advance\@tempdimc\tabcolsep
		\advance\@tempdima1.5\tabcolsep
	\kern-1.5\@tempdimb
	\leaders\vrule
	\hskip\@tempdima\@plus  1fill
	\kern-1.5\@tempdimc
	\hskip-\wd\z@ \@plus -1fill }
\makeatother

\begin{document}

\allowdisplaybreaks

\title{
Background Field Method and
Generalized Field Redefinitions\\
in Effective Field Theories.
}

\date{February 21, 2021}


\author{A. Quadri}
\email{andrea.quadri@mi.infn.it}
\affiliation{INFN, Sezione di Milano, via Celoria 16, I-20133 Milano, Italy}

\begin{abstract}
\noindent
We show that 
in a spontaneously broken  effective gauge field theory, quantized
in a general background $R_\xi$-gauge,
also the background fields undergo a
non-linear (albeit background-gauge invariant)
field redefinition induced by
radiative corrections.
This redefinition proves to be crucial
in order to renormalize the 
coupling constants 
of gauge-invariant operators 
in a gauge-independent way.
The classical background-quantum splitting
is also in general non-linearly deformed (in a non gauge-invariant way) by radiative
corrections. Remarkably, such deformations
vanish in the Landau gauge, to all orders in the
loop expansion.
\end{abstract}


\maketitle

\section{Introduction}

In the absence of direct resonance signals of new physics 
beyond the Standard Model (BSM) at the LHC, 
indirect experimental searches of BSM physics have become increasingly
popular in recent years, e.g.  lepton flavour universality violations ~\cite{Aaij:2014ora,Aaij:2015oid,Sirunyan:2017dhj,Aaboud:2018krd} or searches for
non-resonant Higgs boson pair production (for a recent review see~\cite{Guerrero:2020rlw}).

In this context 
the 
SM Effective Field Theory (SMEFT)~\cite{Buchmuller:1985jz,Grzadkowski:2010es,Brivio:2017vri}
provides a consistent theoretical tool in order
to describe the energy regime up to some
higher energy scale $\Lambda$.
The advantage of the SMEFT is that it
takes into account the constraints arising from
the invariance under the $\rm SU(3) \times SU_L(2) \times U_Y(1)$ gauge group in a model independent way, without the need to know  the precise form of its ultraviolet (UV) completion.

In this approach the SM Lagrangian is supplemented by higher dimensional
gauge-invariant operators suppressed by powers of $\Lambda$.
Renormalizability by power-counting is then lost
and new UV divergences arise order by order in the loop expansion. 
As in any  effective gauge theory, they must be subtracted by 
a combination of generalized (i.e. non-linear and in general
not even polynomial~\cite{Binosi:2019olm}) field redefinitions and the renormalization of the coupling
constants, associated with gauge-invariant operators of increasing dimensions~\cite{Anselmi:2013sx,Gomis:1995jp}. 

That such a program can indeed be completed in a recursive way
by adding local counter-terms  
while preserving the relevant
symmetries of the theory is a key result established many years ago~\cite{Gomis:1995jp} 
in the  setting  of the Batalin-Vilkovisky (BV) formalism (for a review see e.g. \cite{Gomis:1994he}).

The BV formalism can be seen as a generalization of the 
BRST quantization procedure~\cite{Becchi:1974xu,Becchi:1974md,Becchi:1975nq}
that applies also
to non power-counting renormalizable models. The Slavnov-Taylor (ST) identity~\cite{Slavnov:1972fg,Taylor:1971ff},
encoding at the quantum level the 
BRST invariance of the classical gauge-fixed action,
is translated into the BV master equation.

From a physical point of view
the BV master equation ensures, as well as the ST identity,
physical unitarity
of the theory, i.e. the cancellation of unphysical ghosts in the intermediate states~\cite{Becchi:1974xu,Kugo:1977zq,Curci:1976yb,Ferrari:2004pd}.



Due to the huge number of operators 
arising in effective field theories 
it is natural to apply the background field method (BFM)~\cite{DeWitt:1967ub,KlubergStern:1974xv,Boulware:1980av,Hart:1984jy,Abbott:1980hw,Abbott:1983zw,Ichinose:1981uw,Capper:1982tf,Denner:1994xt,Denner:1996wn,Grassi:1999nb,Grassi:1995wr,Becchi:1999ir,Ferrari:2000yp} technique in order to simplify the task
of computing the radiative corrections. The BFM is particularly advantageous since it allows to retain (background) gauge invariance to
all orders in perturbation theory. The resulting
background Ward identity is linear in the quantum
fields, unlike the ST identity, and hence is easier to study.
Use of the BFM has been recently advocated in the context of the (geometric) SMEFT in Refs~\cite{Helset:2018fgq,Corbett:2019cwl,Corbett:2020ymv}.

In power-counting renormalizable theories both the background and the quantum fields
renormalize linearly. 
Linearity of the renormalization of the
background fields together
with background gauge invariance yields
powerful relations between
counter-terms that are one of the
main virtues of the BFM~\cite{Weinberg:1996kr}.

\begin{figure}
    \centering
    \includegraphics[width=5cm, height=5cm]{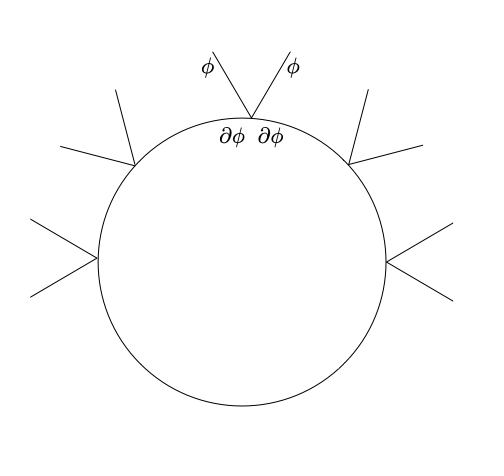}
    \caption{Maximally UV divergent one-loop amplitudes generated by the vertex
    $(\phi\partial\phi)^2$}
    \label{fig:1}
\end{figure}

The situation is significantly more involved in  effective gauge  theories.
For instance a typical  derivative-dependent dim.6 interaction $\sim (\phi \partial \phi)^2$ gives rise already at one loop
to an infinite number of UV-divergent
amplitudes, generated by  
configurations with
two powers of the internal loop momentum 
from the derivative-dependent interaction at each vertex. They are compensated
by two inverse powers from each propagator
(see Figure~\ref{fig:1}), so that
the UV degree of divergence of these Feynman amplitudes
is always $4$, irrespectively of the number of
the external $\phi$-legs.

The task of evaluating the required counter-terms 
in spontaneously broken effective 
gauge field theories is simplified
in the so-called $X$-formalism~\cite{Quadri:2006hr,Quadri:2016wwl,Binosi:2017ubk}
 by the use of a gauge-invariant field coordinate
 for the physical scalar mode, namely
$X_2 \sim  \frac{1}{v} \Big ( \phi^\dagger \phi - \frac{v^2}{2} \Big )$,
where $\phi$ is the usual Higgs doublet
and $v$ its vacuum expectation value (v.e.v.).

Let us consider e.g. the two-derivatives
vertices $(\phi\partial\phi)^2$ arising from
the gauge-invariant interaction
 $\Big (\phi^\dagger \phi - \frac{v^2}{2} \Big) (D^\mu \phi)^\dagger D_\mu \phi \, ,$
$D_\mu$ being the covariant derivative.
This operator is represented in the $X$-formalism
by
$ \sim X_2 (D^\mu \phi)^\dagger D_\mu \phi \,$~\cite{Binosi:2019olm}.
Since the $X_2$-amplitudes are
uniquely fixed by the
functional identities of the theory~\cite{Binosi:2019olm}, one needs
to consider only graphs with internal $X_2$-lines, so that at least one derivative
acts on the external $\phi$-legs, thus reducing
the UV degree of divergence of the amplitudes
that need to be evaluated.
There is only a finite number of
UV divergent amplitudes of this type,
so one can renormalize this sector
of the theory by a finite number of
independent (i.e. not fixed
by the symmetries) local counter-terms, while 
diagrams  in Figure~\ref{fig:1} are automatically
taken into account by an algebraic resummation induced by the
functional identities of the model~\cite{Quadri:2006hr,Quadri:2016wwl,Binosi:2017ubk}.

In effective gauge theories quantum 
fields undergo generalized non-linear
field redefinitions (GFRs).
The $X$-formalism provides 
an effective way  to separate the renormalization
of the gauge coupling constants from
the physically spurious contributions
controlled by the GFRs~\cite{Quadri:2006hr,Quadri:2016wwl,Binosi:2017ubk}.
GFRs play a crucial role in carrying out
the correct recursive
off-shell renormalization of the 
one-particle-irreducible (1-PI) amplitudes,
since only once the appropriate GFRs 
have been implemented, the renormalization
of the coupling constants turns out
to be gauge-independent~\cite{Binosi:2020unh,Binosi:2019olm,Binosi:2019nwz}.

When effective gauge theories
are quantized in the BFM, the question
arises of whether also the background
fields undergo a non-linear
redefinitions, and if such a redefinition
is background gauge-invariant.

This is a non trivial issue that can be
studied by combining the $X$-formalism
with the Algebraic Renormalization
approach to the BFM~\cite{Ferrari:2000yp,Becchi:1999ir,Grassi:1999nb,Grassi:1995wr}.
Compatibilty between
the ST identity and the background Ward
identity is obtained by extending the
BRST differential $s$ to the background fields,
collectively denoted by $\bkg{\Phi}$,
and by pairing them with anticommuting 
variables $\Omega_{\bkg{\Phi}}$, so that
\begin{align}
    s \bkg{\Phi} = \Omega_{\bkg{\Phi}} \, ,
    \qquad s \Omega_{\bkg{\Phi}}  = 0 \, .
\end{align}
The corresponding extended ST identity uniquely
fixes (in a background gauge-invariant
way) the 
dependence of the vertex functional
on the background fields $\bkg{\Phi}$.

In the present paper we extend
the $X$-formalism to the BFM and
study the renormalization of the 
Abelian Higgs-Kibble model
supplemented by dim.6 operators,
as a playground 
towards the renormalization
of the SMEFT in the BFM approach.

We find that:
\begin{enumerate}
   \item the tree-level background-quantum
   splitting
   $$ \Phi = \bkg{\Phi} + Q_\Phi $$
   is in general deformed in a non-linear
    (and gauge-dependent) way, unlike in the power-counting
   renormalizable case where only 
   multiplicative $Z$-factors arise
   both for
   background and quantum fields;
   \item a noticeable exception is 
   the Landau gauge, where no such
   deformation of the tree-level
   background-quantum splitting happens, to all orders in the loop expansion;
   \item as a consequence of
   the radiative corrections
   to the background-quantum splitting and of the GFRs, 
   background fields also
   undergo a non-linear 
   redefinition;
   \item  the redefinition of the background fields is background gauge-invariant. This 
   result follows from non-trivial
   cancellations between the 
   non gauge-invariant contributions to
   the background-quantum splitting
    and the
   non gauge-invariant terms in the GFRs;
    \item  the 
    background and quantum field
    redefinitions are crucial in order
    to properly renormalize the
    coupling constants in a gauge-independent way.
\end{enumerate}

\medskip
The paper is organized as follows.
In Sect.~\ref{sec.1} 
we set up our notations,
present the classical action
of the Abelian Higgs-Kibble model
with dim.6 operators in the
$X$-formalism 
and introduce the BFM tree-level
vertex functional, together
with the background gauge-fixing.
In Sect.~\ref{sec.2} we study the compatibility between
the background Ward identity and the mapping from
the $X$-formalism to
 the standard $\phi$-representation (target theory).
We prove that 1-PI amplitudes in the target theory are background gauge-invariant
if those in the $X$-theory are.
In Sect.~\ref{sec:WI} we study the local solutions to
the background Ward identity that are relevant for
the classification of the UV divergences of the theory.
In Sect.~\ref{sec.4} we solve the
 ST identity
 in order to fix the dependence on the background fields. We find that  
  the tree-level background-quantum splitting is non-trivially deformed
  at the quantum level.
 The gauge dependence of such corrections
 is studied in the Feynman and in the Landau gauge.
 In Sect.~\ref{sec.bgfrs} we obtain
 the generalized field and background
 redefinitions for the $X$-theory 
 by combining the effect of the 
 deformation of the 
 background-quantum splitting and 
 of the GFRs.
 As a non-trivial check, we show that
 at zero quantum fields $Q_\Phi=0$
 subtle cancellations happen 
 that make the vertex functional
 invariant w.r.t. the background
 transformation of the background fields
 only, in agreement with the
 background Ward identity.
 Finally in Sect.~\ref{sec.gfrs.target}
 we provide the explicit form of
 the GFRs both for background and 
 quantum fields in the ordinary
 $\phi$-formalism by applying the mapping
 from the $X$- to the target theory.
 Conclusions are presented in Sect.~\ref{sec.concls}.

 Appendices contain the discussions
 of some aspects of the 
 Algebraic Renormalization of the theory. 
 In Appendix~\ref{app.f.ids}
 we enumerate the functional symmetries of the model.
 Appendix~\ref{app.bkg.gfr} is devoted to the parameterization of
 background gauge field redefinitions. 
 The renormalization of the tadpole and its gauge dependence
are studied in Appendix~\ref{app.tadpole}.

\section{BFM tree-level vertex functional}\label{sec.1}

We start from the tree-level vertex functional  of the Abelian Higgs-Kibble model supplemented by dim.6 gauge-invariant operators  in the so-called $X$-formalism of~\cite{Binosi:2019nwz}:
\begin{align}
	\G^{(0)} & = 
	   \int \!\mathrm{d}^4x \, \Big [ -\frac{1}{4} F^{\mu\nu} F_{\mu\nu} + (D^\mu \phi)^\dagger (D_\mu \phi) - \frac{M^2-m^2}{2} X_2^2 - \frac{m^2}{2v^2} \Big ( \phi^\dagger \phi - \frac{v^2}{2} \Big )^2 \nonumber \\
	& - \bar c (\square + m^2) c + \frac{1}{v} (X_1 + X_2) (\square + m^2) \Big ( \phi^\dagger \phi - \frac{v^2}{2} - v X_2 \Big ) \nonumber \\
	&  +\frac{z}{2} \partial^\mu X_2 \partial_\mu X_2 + \frac{g_1 v}{\Lambda^2} X_2 (D^\mu \phi)^\dagger (D_\mu \phi) 
	+  \frac{g_2 v}{\Lambda^2} X_2 F_{\mu\nu}^2 
	+  \frac{g_3 v^3 }{6 \Lambda^2} X_2^3 
	\nonumber \\
	& + \T(D^\mu \phi)^\dagger (D_\mu \phi) +
	U F_{\mu\nu}^2 + R X_2^2
	\nonumber \\
	& + \frac{\xi b^2}{2} -  b \Big ( \partial A + \xi e v \chi \Big ) + \bar{\omega}\Big ( \square \omega + \xi e^2 v (\sigma + v) \omega\Big ) \nonumber \\
	&  + \bar c^* \Big ( \phi^\dagger \phi - \frac{v^2}{2} - v X_2 \Big ) + \sigma^* (-e \omega \chi) + \chi^* e \omega (\sigma + v) \Big ].
	\label{tree.level}
\end{align}
The field content of the model includes the Abelian gauge field
$A_\mu$,  the usual scalar field 
$$\phi \equiv \frac{1}{\sqrt{2}} (\phi_0 + i \chi) = \frac{1}{\sqrt{2}} (\sigma + v + i \chi)\, , \qquad \phi_0 = \sigma + v \, ,$$ 
with $v$ denoting the v.e.v., and a singlet field $X_2$ describing 
in a gauge invariant way the physical scalar mode of mass $M$.
Indeed, if one goes on-shell in Eq.(\ref{tree.level}) with the auxiliary field $X_1$, that plays the role of a Lagrange multiplier, we obtain the constraint 
$$
    (\square + m^2) \Big (
    \phi^\dagger \phi - \frac{v^2}{2} - v X_2 \Big ) = 0,
$$
so that the field $X_2$ must fulfill the condition
$X_2 = \frac{1}{v} \Big (
    \phi^\dagger \phi - \frac{v^2}{2} \Big ) + \eta,$ $\eta$
being a scalar field of mass $m$. 
However, it can be proven that in perturbation theory 
the correlators of the mode $\eta$ with any gauge-invariant operators vanish~\cite{Binosi:2019olm}, so that one can safely set $\eta =0$
and by going on-shell perform  in Eq.~(\ref{tree.level}) the substitution
$$X_2 \sim \frac{1}{v} ( \phi^\dagger \phi - v^2/2) \, .$$
The $m^2$-term cancels out and one gets back the usual Higgs quartic potential with coefficient $\sim M^2/2v^2$ plus the set of dim.6 parity-preserving operators arising from 
the third line of Eq.(\ref{tree.level}) (we use the same notations 
as in~\cite{Binosi:2020unh}):
\begin{subequations}
\begin{align}
    \op{1}^{[6]} &= \int \d  \, F_{\mu\nu}^2 \Big ( \phi^\dagger \phi - \frac{v^2}{2} \Big ) \sim \int \d  \,  v X_2 F_{\mu\nu}^2 ,\\
    \op{2}^{[6]} &= \int \d  \,  \Big ( \phi^\dagger \phi - \frac{v^2}{2} \Big )^3 \sim \int \d  \, v^3 X_2^3, \\
    \op{3}^{[6]} &= \int \d  \,  \Big ( \phi^\dagger \phi - \frac{v^2}{2} \Big ) \square \Big ( \phi^\dagger \phi - \frac{v^2}{2} \Big )
    \sim \int \d  \, v^2 X_2 \square X_2 , \\
    \op{4}^{[6]} &= \int \d  \,  \Big ( \phi^\dagger \phi - \frac{v^2}{2} \Big ) (D^\mu \phi)^\dagger D_\mu \phi \sim  \int \d  \, v X_2 (D^\mu \phi)^\dagger D_\mu \phi.
\end{align}
\end{subequations}

We notice that the parameter $m^2$ must disappear in the correlators of the gauge-invariant operators of the target theory (i.e. the one obtained
by going on-shell with the $X_{1,2}$-fields), as can be checked explicitly at the one loop order~\cite{Binosi:2019olm,Binosi:2020unh,Binosi:2019nwz,Binosi:2017ubk}.

In Eq.(\ref{tree.level}) $\bar \omega, \omega$  are the 
Faddeev-Popov antighost and ghost fields, while $b$ is the
Nakanishi-Lautrup field enforcing the gauge-fixing condition
$$ {\cal F}_\xi = 0 $$
with
\begin{align}
    {\cal F}_\xi \equiv \partial A + \xi e v\chi \, ,
    \label{quantum.g.f}
\end{align}
$\xi$ being the gauge parameter.

The tree-level vertex functional (\ref{tree.level}) is invariant both under the usual
{\em gauge} BRST symmetry 
\begin{align}
& s A_\mu = \partial_\mu \omega \, ; \quad s\phi = i e \omega \phi \, ; \quad s \sigma = -e \omega \chi ; \quad 
s \chi = e \omega (\sigma + v) ; \quad s \omega = 0 \, ; 
\nonumber \\
& s \bar c = b \, ; \quad s b= 0 \, ; 
\quad  s X_1 = s X_2 = s c = s \bar c = 0 \, , 
\end{align}
and 
a {\em constraint} BRST symmetry
\begin{align}
	\s X_1 = v c; \, \quad \s c  = 0; \quad \s \bar c = \phi^\dagger\phi - \frac{v^2}{2} - v X_2 \, ,
	\label{u1.brst}
\end{align}
while all other fields are $\s$-invariant.
The latter symmetry ensures that the number of physical degrees of freedom in the scalar sector remains unchanged in the $X$-formalism with respect to the standard formulation relying only on the field $\phi$~\cite{Quadri:2006hr,Quadri:2016wwl}.
$c, \bar c$ are the ghost and antighost fields of the
constraint BRST symmetry. They are free. 

The two BRST differentials $s,\s$ anticommute.

Several external sources
need to be introduced in the vertex functional
(\ref{tree.level}) in order to formulate at the quantum level the symmetries of the theory, as a consequence of the 
non-linearity in the quantized fields of the operators involved: the antifields~\cite{Gomis:1994he} $\sigma^*, \chi^*$,
{\it i.e.}, the external sources coupled to the relevant BRST transformations  that are non-linear in the quantized fields, the antifield $\bar c^*$ coupled to the constraint
BRST variation of $\bar c$ in Eq.(\ref{u1.brst}), 
and the sources $T_1, U$ and $R$, coupled
to the gauge-invariant operators in the fourth line of 
Eq.(\ref{tree.level}). The latter sources are needed in order to
define  at the quantum level the $X_{1,2}$-equations of motion, as summarized in Appendix~\ref{app.f.ids}.

The main virtue of this approach is that several relations 
among 1-PI Green's functions of the effective field theory, that 
are hidden in the standard formulation, becomes manifest 
as they are encoded in the $X_{1,2}$-equation and the related
system of external sources.
In particular the $X$-formalism is  suited in order to evaluate the GFRs and
disentangle the gauge-invariant
renormalization of the 
coupling constants~\cite{Binosi:2019olm,Binosi:2019nwz,Binosi:2020unh}.

In order to formulate the theory in the background field method we
introduce the background gauge field $\bkg{A}_\mu$
and the background scalar $\bkg{\phi} \equiv \frac{1}{\sqrt{2}}
( \bkg{\sigma} + v + i \bkg{\chi})$.
They transform as the corresponding fields under a background gauge transformation of parameter $\alpha$, namely
\begin{align}
    & \delta A_\mu = \partial_\mu \alpha \, , \quad
      \delta \bkg{A}_\mu = \partial_\mu \alpha \, , \quad 
      \delta \phi = i e \alpha \phi \, , \quad
      \delta \bkg{\phi} = i e \alpha \bkg{\phi} \, .
\end{align}
If the gauge-fixing functional ${\cal F}_\xi$ in Eq.(\ref{quantum.g.f}) is replaced by
\begin{align}
    \bkg{\cal F}_\xi = \partial^\mu (A_\mu - \bkg{A}_\mu) + ~ \xi e (\bkg{\phi}_0 \chi - \bkg{\chi}\phi_0) \, , 
    \label{bkg.g.f}
\end{align}
the tree-level vertex functional $\G^{(0)}$ becomes background gauge invariant provided that: i) all other fields and external sources are
required to be $\delta$-invariant, with the exception of the antifields $\sigma^*,\chi^*$;
ii) $\sigma^*,\chi^*$ are gathered in a complex antifield $\phi^* \equiv \frac{1}{\sqrt{2}} ( \sigma^* + i \chi^*)$ transforming as a scalar in the fundamental representation:
\begin{align}
    \delta \phi^* = i e \alpha \phi^* \, .
\end{align}
In order to ensure the compatibility of the background gauge invariance with the ST identity, one also needs to introduce for each
background field $\bkg{\Phi}$ an anti-commuting variable 
$\Omega_{\Phi}$ pairing with the background field into a BRST doublet~\cite{Grassi:1999nb,Becchi:1999ir,Ferrari:2000yp}:
\begin{align}
    s \bkg{A}_\mu = \Omega_\mu \, , \quad
    s \bkg{\sigma} = \Omega_{\hat\sigma} \, , \quad
    s \bkg{\chi} = \Omega_{\hat \chi} \, ,
    \quad s \Omega_\mu =
    s \Omega_{\hat\sigma} = \Omega_{\hat \chi} = 0 \, .
\end{align}
This procedure uniquely fixes  the dependence of the vertex
functional on the background fields in the sector at zero ghost number, since in this sector the background-dependent part of the vertex functional can be recovered by a canonical transformation that respects the ST identity (when the latter is 
equivalently rewritten as the Batalin-Vilkovisky master equation)~\cite{Binosi:2012st,Anselmi:2013kba}. Being a canonical transformation, the physical content of the theory is not modified by the introduction of the background fields~\cite{Grassi:1995wr,Becchi:1999ir,Ferrari:2000yp,Binosi:2012st,Anselmi:2013kba}.

The tree-level vertex functional 
$\Gamma^{(0)}$
in the presence of the
background fields thus
 acquires an $\Omega$-dependence generated by 
the gauge-fixing term:
\begin{align}
    \G^{(0)}_{\rm g.f.} & = 
    \int \d s \Big [ \bar \omega \Big ( 
          \xi \frac{b}{2} - \bkg{\cal F}_\xi \Big ) \Big ] 
          \nonumber \\
          & =\int \d \Big [ 
          \frac{\xi b^2}{2} - b \Big ( \partial A + \xi e v \chi \Big ) 
          + \bar \omega \Big ( \square \omega + \xi e^2 v (\sigma + v) \omega \Big )
          \nonumber \\
          & \qquad \qquad + b \Big ( \partial \bkg{A} - \xi e \bkg{\sigma} \chi +  \xi e \bkg{\chi} (\sigma + v) \Big ) + \bar \omega \xi e^2 \Big ( \bkg{\sigma} (\sigma + v) + \bkg{\chi} \chi  \Big ) \omega \nonumber \\
          & \qquad \qquad - \bar \omega \partial^\mu \Omega_\mu
          + \xi e \bar \omega \Omega_{\bkg{\sigma}} \chi
          - \bar \omega \xi e \Omega_{\bkg{\chi}} (\sigma + v)
          \Big ] \ .
\label{bkg.gf}
\end{align}
The last two lines in the above equation contain the additional
terms proportional to the background fields and their
BRST partners. At $\bkg{A}_\mu = \bkg{\sigma} = \bkg{\chi} = 0$
as well as $\Omega_{\bkg{\sigma}} = \Omega_{\bkg{\chi}} = 
\Omega_{\mu} = 0$ we recover the gauge-fixing
and ghost terms in Eq.(\ref{tree.level}).

The background tree-level vertex functional is then 
obtained by the replacement
\begin{align}
\G^{(0)} & \rightarrow \G^{(0)} + \int \d \Big [
     b \Big ( \partial \bkg{A} - \xi e \bkg{\sigma} \chi +  \xi e \bkg{\chi} (\sigma + v) \Big ) 
      + \bar \omega \xi e^2 \Big ( \bkg{\sigma} (\sigma + v) + \bkg{\chi} \chi  \Big ) \omega  \nonumber \\
     & \qquad \qquad \qquad \quad
     - \bar \omega \partial^\mu \Omega_\mu
     + \xi e \bar \omega \Omega_{\bkg{\sigma}} \chi
          -  \xi e \bar \omega \Omega_{\bkg{\chi}} (\sigma + v)
          \Big ] \, .
    \label{bfm.tl.vf}
\end{align}

The ghost number is assigned as follows.
$A_\mu, \sigma, \chi, X_1, X_2, b$ have ghost number zero.
$c,\omega$ and the background BRST partners $\Omega_\mu,
\Omega_{\bkg{\sigma}}, \Omega_{\bkg{\chi}}$ have ghost number one.
$\bar c^*$ has ghost number zero. The antifields $\sigma^*, \chi^*$
have ghost number -1.

Since the theory is non-anomalous, the full vertex functional
$\Gamma$ is invariant under the ST identity, the background Ward identity and the $X_{1,2}$-equations, as summarized in 
Appendix~\ref{app.f.ids}. 

$\Gamma$ can be expanded
in the loop parameter as follows:
\begin{align}
    \Gamma = \sum_{n=0}^\infty \hbar^n \Gamma^{(n)} \, .
\end{align}
Perturbation theory is carried out order by order in the loop expansion by recursively imposing the functional identities
of the model while subtracting the UV divergences 
by means of suitable local (in the sense of formal power series)
counter-terms. 

\section{Background Ward
identity for the target theory}\label{sec.2}


Eventually we are interested in the 
1-PI Green's functions
in the standard
$\phi$-formalism, that
are obtained
from the vertex functional
of the $X$-theory
by going on shell
w.r.t. the fields $X_{1,2}$~\cite{Binosi:2020unh,Binosi:2019olm,Binosi:2019nwz,Binosi:2017ubk}.

The procedure amounts to carry out the replacements
in Eq.(\ref{X12.sols}) and then substitute $X_{1,2}$ with
the solution of their equations of motion, order by order
in the loop expansion.

Let us denote by $\widetilde{\G}$ the 
vertex functional of the target theory 
(i.e. the generating functional of the 1-PI Green's functions
in the $\phi$-formalism).
Since the functional differential operators 
for the $X_{1,2}$-equations in Eqs.(\ref{X1.eq}) and
(\ref{X2.eq}) and for the background Ward identity
in Eq.(\ref{bkg.wi}) commute:
\begin{align}
    [ {\cal B}_{X_1}, {\cal W}] =    [ {\cal B}_{X_2}, {\cal W}] = 0 \, ,
\end{align}
we conclude that $\widetilde{\G}$   is also background
gauge-invariant.

It is instructive to check this result at one loop order
in the sector of operators
up to dimension $6$, for which
the explicit form of the mapping
has been worked out 
in~\cite{Binosi:2019nwz,Binosi:2019olm,Binosi:2020unh}.
At one loop we need to solve the tree-level equations of motion
for $X_{1,2}$~\cite{Binosi:2020unh}.
The $X_1$-equation of motion yields
\begin{align}
    X_2 = \frac{1}{v} \Big ( \phi^\dagger \phi - \frac{v^2}{2} \Big ),
    \label{tl.X1.eq}
\end{align}
while the classical $X_2$-equation of motion gives (at zero external sources)
\begin{align}
    (\square + m^2)(X_1 + X_2) & = - (M^2 - m^2) X_2 - z \square X_2 + \frac{g_1 v}{ \Lambda^2} (D^\mu \phi)^\dagger D_\mu \phi + 
    \frac{g_2 v}{ \Lambda^2} F_{\mu\nu}^2 + \frac{g_3 v^3}{2  \Lambda^2} X_2^2.
    \label{tl.X2.eq}
\end{align}
By inserting Eqs.~(\ref{tl.X1.eq}) and~(\ref{tl.X2.eq}) into the replacements
in Eq.(\ref{X12.sols}) we obtain the explicit form of the mapping 
at one loop:
\begin{subequations}
\begin{align}
 \tbarc \rightarrow &
 - \frac{(M^2 - m^2)}{v^2} \Big ( \phi^\dagger \phi - \frac{v^2}{2} \Big )  - \frac{z}{v^2} \square \Big ( \phi^\dagger \phi - \frac{v^2}{2} \Big ) + \frac{g_1}{ \Lambda^2} (D^\mu \phi)^\dagger D_\mu \phi + 
    \frac{g_2}{ \Lambda^2} F_{\mu\nu}^2\nonumber \\
    & + \frac{g_3}{2  \Lambda^2} \Big ( \phi^\dagger \phi - \frac{v^2}{2} \Big )^2,
    \label{mapping1}\\
    \tT \rightarrow &\frac{g_1 }{\Lambda^2} \Big ( \phi^\dagger \phi - \frac{v^2}{2} \Big );\qquad 
    \tU \rightarrow \frac{g_2}{\Lambda^2}  \Big ( \phi^\dagger \phi - \frac{v^2}{2} \Big );\qquad
    \tR \rightarrow  \frac{g_3 v^2}{2 \Lambda^2} \Big ( \phi^\dagger \phi - \frac{v^2}{2} \Big ). 
\label{mapping2}
\end{align}
\label{mapping}
\end{subequations}

It can be seen by direct inspection that the r.h.s. of the above Equations only contain gauge-invariant operators.
Since $\G^{(1)}$ is background gauge invariant and the replacements
in Eqs.(\ref{mapping1}) and (\ref{mapping2}) transform
background gauge-invariant sources into background gauge-invariants
combinations in the target theory, we conclude that 
$\widetilde{\G}^{(1)}$ is 
automatically background gauge-invariant.

\section{Local solutions to the background Ward identity}\label{sec:WI}

Let us denote by $\overline{\G}^{(n)}$ the UV-divergent part of the $n$-th order vertex functional.
Provided that the UV divergences have been subtracted up to order $n-1$ in a way to preserve the symmetries of the theory, 
$\overline{\G}^{(n)}$ is a local functional (in the sense of
formal power series) in the fields, the external sources and 
their derivatives.

If the regularization scheme is symmetric (as it happens
e.g. for dimensional regularization), 
UV divergences must also fulfill the same background Ward identity
in Eq.(\ref{bkg.wi}):
\begin{align}
    {\cal W} ( \overline{\G}^{(n)} ) = 0 \, .
    \label{n.th.order.WI}
\end{align}
Since the $n$-th order UV divergences are local,
we need to solve Eq.(\ref{n.th.order.WI}) in the space of local functionals. Moroever by Eq.(\ref{redef.chistar}) we can use the redefined antifield ${\chi^{*}}^\prime$ and then set $\bar \omega=b=0$ (since
the $n$-th order vertex functional $n\geq 1$ does not
depend on $b$ and the only dependence on the antighost
$\omega$ is via ${\chi^{*}}^\prime$,
as can be seen from Eq.(\ref{n.ag})).

An efficient way to obtain the most general solution to 
Eq.(\ref{n.th.order.WI}) in this 
functional space 
is to carry out the following change of variables:
\begin{gather}
    A_\mu \rightarrow Q_\mu = A_\mu - \bkg{A}_\mu \, , \quad
    \sigma \rightarrow \tilde{\phi}_0 - v \equiv 
    \frac{1}{v} ( \bkg{\phi}_0 \phi_0 + \bkg{\chi} \chi) - v 
    \, , \quad
    \chi \rightarrow \tilde{\chi} \equiv \frac{1}{v}
    ( \bkg{\phi}_0 \chi - \phi_0 \bkg{\chi} ) \, , 
    \nonumber \\
    \sigma^* \rightarrow \tilde \sigma^* \equiv \frac{1}{v} ( \bkg{\phi}_0 \sigma^* + \bkg{\chi} \chi^*) \, ,
    \quad
    {\chi^{*}}^\prime \rightarrow \widetilde{ {\chi^{*}}^\prime} \equiv \frac{1}{v} ( \bkg{\phi}_0 {\chi^{*}}^\prime - \sigma^* \bkg{\chi} ) \, .
    \label{bleached.bkg}
\end{gather}
It is easy to see that $Q_\mu, \tilde{\phi}_0,\tilde \chi, \tilde \sigma^*, \widetilde{ {\chi^{*}}^\prime}$ are gauge-invariant. 
Moreover they reduce to the original fields and antifields
at zero backgrounds.

Accordingly,
the most general solution 
$\overline{\G}^{(n)}[A_\mu, \sigma, \chi; \bkg{A}_\mu, \bkg{\sigma}, \bkg{\chi};
\sigma^*, \chi^*]$
to Eq.(\ref{n.th.order.WI}) 
can be written as follows:
\begin{align}
    \overline{\G}^{(n)} 
    = \overline{\G}^{(n)}[Q_\mu, \tilde{\phi}_0 - v, \tilde{\chi};0,0,0;\tilde \sigma^*,\tilde \chi^*] + {\cal P}^{(n)}
    (\bkg{A}_\mu,\bkg{\sigma},\bkg{\chi}) 
    \, . 
    \label{bkg.gen.sol}
\end{align}
The first term in the r.h.s. of Eq.(\ref{bkg.gen.sol}) is obtained by replacing the quantum fields
$(A_\mu, \sigma, \chi)$ 
and their antifields
with their background gauge-invariant counterparts in Eq.(\ref{bleached.bkg}).
The second term ${\cal P}^{(n)}$ is the most general solution in the kernel of the operator ${\cal W}$, namely a gauge invariant 
formal power series built out from the
background fields
scalar $\bkg{\phi}$ and its background covariant derivatives
and from the background field strength $\bkg{F}_{\mu\nu} = \partial_\mu \bkg{A}_\nu - \partial_\nu \bkg{A}_\mu$ and its
ordinary derivatives, that vanishes at zero background fields.

The background Ward identity is unable to fix the ambiguities
encoded by ${\cal P}^{(n)}(\bkg{A}_\mu,\bkg{\sigma},\bkg{\chi})$.
One thus needs to make recourse to the extended ST identity in order
to select out of the general solution
to the background Ward identity in Eq.(\ref{bkg.gen.sol})
the unique vertex functional (in the sector with
zero ghost number) depending on the background fields
and compatible with the ST identity itself.

A remark is in order here. 
A different basis is often used in BFM
calculations, namely
the quantum fields are defined as
$Q_\mu \equiv A_\mu - \bkg{A}_\mu, q_\sigma \equiv \sigma - \bkg{\sigma}, q_\chi \equiv \chi - \bkg{\chi}$,
i.e. the variables over which
one integrates in the path integral.
We collectively denote these fields
by $Q_\Phi$.

The background Ward identity in 
the $Q_\Phi$-variables 
reads
\begin{align}
    {\cal W} (\Gamma ) = &
        - e q_\chi \frac{\delta \G}{\delta q_\sigma} 
    + e q_\sigma  \frac{\delta \G}{\delta q_\chi} 
      - \partial^\mu \frac{\delta \G}{\delta \bkg{A}^\mu}
    - e \bkg{\chi} \frac{\delta \G}{\delta \bkg{\sigma}} 
    + e (\bkg{\sigma} + v) \frac{\delta \G}{\delta \bkg{\chi}} \nonumber \\
    & 
     - e \chi^* \frac{\delta \G}{\delta \bkg{\sigma^*}} 
    + e \sigma^* \frac{\delta \G}{\delta \chi^*} = 0 \, .
    \label{bkg.wi.qnt}
\end{align}
We notice that at $Q_\Phi=0$
Eq.(\ref{bkg.wi.qnt}) states that
the vertex functional at zero quantum fields
is background-gauge invariant:
\begin{align}
    {\cal W} (\left . \Gamma \right |_{Q_\Phi=0}) = & \Big \{ 
      - \partial^\mu \frac{\delta}{\delta \bkg{A}^\mu}
    - e \bkg{\chi} \frac{\delta}{\delta \bkg{\sigma}} 
    + e (\bkg{\sigma} + v) \frac{\delta}{\delta \bkg{\chi}} 
     - e \chi^* \frac{\delta}{\delta \bkg{\sigma^*}} 
    + e \sigma^* \frac{\delta}{\delta \chi^*}
    \Big \} \left . \Gamma \right |_{Q_\Phi=0}
    = 0 \, .
    \label{bkg.wi.zero.qnt}
\end{align}
At $Q_\Phi=0$
the redefined fields 
in Eq.(\ref{bleached.bkg})
reduce to gauge-invariant combinations, namely
\begin{align}
    \left . Q_\mu \right |_{Q_\Phi = 0} = 0 \, , \quad
    \left . \tilde{\phi}_0  \right |_{Q_\Phi = 0} = 
    \frac{1}{v} ( \bkg{\phi}_0^2 + \bkg{\chi}^2 )  \, , \quad
    \left . \tilde{\chi} \right |_{Q_\Phi = 0} = 0 \, . 
    \label{zero.qf.limit}
\end{align}
As a consequence of Eq.(\ref{zero.qf.limit}), at zero quantum fields 
$\overline{\G}^{(n)}$ in Eq.(\ref{bkg.gen.sol})
reduces to a background gauge-invariant functional, in agreement with
Eq.(\ref{bkg.wi.zero.qnt}).

\section{Background-quantum splitting}
\label{sec.4}

In the physical sector at zero ghost number the dependence
on the background fields is uniquely fixed by the ST identity
in Eq.(\ref{sti}) once the 1-PI Green's functions of the quantized
fields and the correlators involving the sources $\Omega_{\bkg{\Phi}}$
are known.



By taking a derivative w.r.t $\Omega_\mu, \Omega_{\bkg{\sigma}}, \Omega_{\bkg{\chi}}$ and then 
setting $c=\omega=\Omega_\mu=\Omega_{\bkg{\sigma}}=\Omega_{\bkg{\chi}} = b= 0$ we get
\begin{align}
& \G'_{\bkg{\sigma}} = - \int \d 
\Big [
\G'_{\Omega_{\bkg{\sigma}} \sigma^*} \G'_{\sigma} +
\G'_{\Omega_{\bkg{\sigma}} \chi^*} \G'_{\chi} \Big ] \, , \nonumber \\
& \G'_{\bkg{\chi}} = - \int \d 
\Big [
\G'_{\Omega_{\bkg{\chi}} \sigma^*} \G'_{\sigma} +
\G'_{\Omega_{\bkg{\chi}} \chi^*} \G'_{\chi} \Big ] \, .
\label{bkg.eoms}
\end{align}
In the above equation we have denoted by a prime the 
functionals evaluated at 
$c=\omega=\Omega_\mu=\Omega_{\bkg{\sigma}}=\Omega_{\bkg{\chi}} = b = 0$.
Moreover 
in order to simplify the notations
we  denote by a subscript the functional differentiation w.r.t 
the field or external source, e.g. $\G_\chi = \frac{\delta \G}{\delta \chi}$.
When the momenta of the fields and external sources
are displayed as arguments of the corresponding amplitudes,
we understand that the functional derivatives 
of the vertex functional are evaluated
at zero fields and external sources, i.e. we refer to
the specific 1-PI amplitudes.
For instance the two-point 1-PI function with one $\sigma$ 
and one background $\bkg{\sigma}$ legs will be denoted
by $\G^{(1)}_{\bkg{\sigma}(-p) \sigma(p)}.$

If one would use  $q_\sigma, q_\chi$ with the corresponding
antifields $q^*_\sigma, q^*_\chi$, an extra dependence
on $\Omega_{\bkg{\sigma}}, \Omega_{\bkg{\chi}}$ would arise,
since
\begin{align}
 &  s q_\sigma = s \sigma - s \bkg{\sigma} = 
          - e \omega \chi - \Omega_{\bkg{\sigma}} \ , \quad
 & s q_\chi = s \chi - s \bkg{\chi} = 
 e \omega (\sigma + v) - \Omega_{\bkg{\chi}} \, ,
\end{align}
so that $\G^{(0)}_{q^*_\sigma \Omega_{\bkg{\sigma}}}$
and $\G^{(0)}_{q^*_\chi \Omega_{\bkg{\chi}}}$ would not vanish,
thus introducing additional terms in the r.h.s. of Eq.(\ref{bkg.eoms}).
For this reason we prefer to use the $(A_\mu,\sigma,\chi)$-basis
in solving the extended ST identity
for the background dependence.

Let us project Eq.(\ref{bkg.eoms}) 
at first order in the loop expansion.
Since we use the basis
$(A_\mu,\sigma,\chi)$,
there is no tree-level
1-PI amplitude involving
the antifields $\sigma^*,\chi^*$
together with the background ghosts
$\Omega_{\bkg{\sigma}},\Omega_{\bkg{\chi}}$.
Hence we find
\begin{align}
& \G^{(1)'}_{\bkg{\sigma}} = - \int \d 
\Big [
\G^{(1)'}_{\Omega_{\bkg{\sigma}} \sigma^*} \G^{(0)'}_{\sigma} +
\G^{(1)'}_{\Omega_{\bkg{\sigma}} \chi^*} \G^{(0)'}_{\chi} \Big ] \, , \nonumber \\
& \G^{(1)'}_{\bkg{\chi}} = - \int \d 
\Big [
\G^{(1)'}_{\Omega_{\bkg{\chi}} \sigma^*} \G^{(0)'}_{\sigma} +
\G^{(1)'}_{\Omega_{\bkg{\chi}} \chi^*} \G^{(0)'}_{\chi} \Big ] \, .
\label{bkg.eoms.1loop}
\end{align}
We are interested in the background dependence of the UV divergences of the theory, that are local. Hence we can 
solve Eqs.(\ref{bkg.eoms.1loop}) in the space of local functionals.
Moreover at $b=0$
there is no dependence of the
tree-level vertex functional on the background, again as a consequence of the use
of the basis $(A_\mu,\sigma,\chi)$.

Thus in order to recover the full dependence on the background fields, once one knows the amplitudes at zero background,
we just need to expand the
kernels $\G^{(1)'}_{\Omega_{\bkg{\sigma}} \sigma^*},\G^{(1)'}_{\Omega_{\bkg{\sigma}} \chi^*},\G^{(1)'}_{\Omega_{\bkg{\chi}} \sigma^*}, \G^{(1)'}_{\Omega_{\bkg{\chi}} \chi^*}$
in powers of the fields, antifields and the backgrounds and then
solve the functional differential equations Eq.(\ref{bkg.eoms.1loop})
by integrating over $\hat\sigma, \hat\chi$.

Since the kernels are gauge-dependent,  we proceed to a separate discussion
for the Feynman and the Landau gauge.

\subsection{Feynman gauge}

In the Feynman gauge $\xi=1$ the kernels are non-vanishing.
We notice that 
by power-counting they  contain at most logarithmic divergences, so we can drop derivative-dependent terms in their local expansion 
around zero momentum and write (we omit
the coefficients vanishing by parity):
\begin{align}
  \overline{\G}^{(1)'}_{\Omega_{\bkg{\sigma}} \sigma^*} =  \int \d \Big [ &
 \gamma_{\Omega_{\bkg{\sigma}} \sigma^*} + 
  \gamma_{\Omega_{\bkg{\sigma}} \sigma^* \sigma} \sigma +
  \gamma_{\Omega_{\bkg{\sigma}} \sigma^* \bkg{\sigma}} \bkg{\sigma} 
  + \frac{1}{2} \gamma_{\Omega_{\bkg{\sigma}} \sigma^* \sigma\sigma} \sigma^2
 +  \frac{1}{2} \gamma_{\Omega_{\bkg{\sigma}} \sigma^* \chi\chi} \chi^2 
\nonumber \\
  & 
 + \gamma_{\Omega_{\bkg{\sigma}} \sigma^* \sigma\bkg{\sigma}} \sigma\bkg{\sigma} 
  + \gamma_{\Omega_{\bkg{\sigma}} \sigma^* \chi\bkg{\chi}} \chi\bkg{\chi} 
+ \frac{1}{2} \gamma_{\Omega_{\bkg{\sigma}} \sigma^* \bkg{\sigma}\bkg{\sigma}} \bkg{\sigma}^2
 +  \frac{1}{2} \gamma_{\Omega_{\bkg{\sigma}} \sigma^* \bkg{\chi}\bkg{\chi}} \bkg{\chi}^2 +
  \gamma_{\Omega_{\bkg{\sigma}} \sigma^* T_1} T_1 + \dots 
 \Big ] \, .
 \label{exp.kernel}
\end{align}
The dots stand for terms with more than two fields $\sigma,\chi$ and their backgrounds 
as well as additional powers of the external sources.
We truncate the expansion to the order required
for the comparison with the explicit results 
of~\cite{Binosi:2020unh}.

More specifically the coefficients can be obtained 
by evaluating the UV divergent part of the 1-PI Green's functions involving insertions
of  $\Omega_{\bkg{\sigma}}, \sigma^*$ and the other
fields and external sources, so for instance
\begin{align}
    & \gamma_{\Omega_{\bkg{\sigma}} \sigma^*} = \left . 
     \overline{\G}^{(1)}_{\Omega_{\bkg{\sigma}(-p)} \sigma^*(p)} 
    \right |_{p=0} \, , 
    & 
     \gamma_{\Omega_{\bkg{\sigma}} \sigma^* \sigma} = 
    \left . \overline{\G}^{(1)}_{\Omega_{\bkg{\sigma}(-p_1-p_2)} \sigma^*(p_1) \sigma(p_2)} \right |_{p_1=p_2=0} \, , 
\end{align}
and so on.
A similar expansion holds for the other kernels.

By explicit computation we find the following
results (to the accuracy required to renormalize
dim.6 operators~\cite{Binosi:2020unh})
\begin{align}
\overline{\G}^{(1)'}_{\Omega_{\bkg{\sigma}} \sigma^*}  
& = \int \d \frac{M_A^2}{8 \pi^2 v^2}\frac{1}{\epsilon}
\Big [ 1 - \frac{z}{1+z} \frac{\chi^2}{v^2} - T_1
+ \dots \Big ] \, , \nonumber \\
\overline{\G}^{(1)'}_{\Omega_{\bkg{\sigma}} \chi^*}  
& = \int \d \frac{M_A^2}{8 \pi^2 v^2}\frac{1}{\epsilon}
\frac{\chi}{v} \Big [ \frac{z}{1+z} -
\frac{z(z-1)}{(1+z)^2}\frac{\sigma}{v}+ \dots \Big ] \, ,
\nonumber \\
\overline{\G}^{(1)'}_{\Omega_{\bkg{\chi}} \sigma^*}  
& = \int \d \frac{M_A^2}{8 \pi^2 v^2}\frac{1}{\epsilon}
\frac{\chi}{v} \Big [\frac{z}{1+z} - \frac{z(z-1)}{(1+z)^2}\frac{\sigma}{v} + \dots
\Big ] \, ,
\nonumber \\
\overline{\G}^{(1)'}_{\Omega_{\bkg{\chi}} \chi^*}  
& = \int \d \frac{M_A^2}{8 \pi^2 v^2}\frac{1}{\epsilon}
\Big [\frac{z}{1+z} - \frac{2z}{(1+z)^2}\frac{\sigma}{v} +
\frac{z(3z-1)}{(1+z)^3}\frac{\sigma^2}{v^2}
+ \frac{z^2}{(1+z)^2}\frac{\chi^2}{v^2}  \nonumber \\
& \qquad \qquad
- \frac{1}{(1+z)^2} T_1 +
\dots \Big ] \, .
\label{kernels}
\end{align}
We notice that in this specific
case there is no
dependence of the kernels in
Eq.(\ref{kernels}) on the background fields, 
so the integration of Eq.(\ref{bkg.eoms.1loop})   is trivial and yields
a linear dependence on the background fields themselves:
\begin{align}
 \overline{\G}^{(1)'} = - \int \d \Big [
 \Big ( \bkg{\sigma} ~ \overline{\G}^{(1)'}_{\Omega_{\bkg{\sigma}} \sigma^*} + \bkg{\chi} ~
 \overline{\G}^{(1)'}_{\Omega_{\bkg{\chi}} \sigma^*} 
 \Big ) \G^{(0)'}_{\sigma} + 
 \Big ( \bkg{\sigma} ~ \overline{\G}^{(1)'}_{\Omega_{\bkg{\sigma}} \chi^*} + \bkg{\chi}~ \overline{\G}^{(1)'}_{\Omega_{\bkg{\chi}} \chi^*} 
 \Big ) \G^{(0)'}_{\chi} \Big ] + 
 \left . \overline{\G}^{(1)'} \right |_{\bkg{\sigma} = \bkg{\chi} = 0} \, .
 \label{bkg.1loop.feyn}
\end{align}
%

%

The last term in Eq.(\ref{bkg.1loop.feyn})  
denotes the UV divergent part of the
vertex functional at zero background fields. 
It has been evaluated in~\cite{Binosi:2020unh} 
for the relevant sector of operators up to dimension $6$.

Eq.(\ref{bkg.1loop.feyn}) is of particular significance.
It states that the background-quantum splitting is non-trivially
modified at the quantum level according to the following redefinitions:
\begin{align}
    & \sigma 
    \rightarrow \bkg{\sigma} + q_\sigma 
      -  \bkg{\sigma} ~ \overline{\G}^{(1)'}_{\Omega_{\bkg{\sigma}} \sigma^*}- \bkg{\chi} ~
 \overline{\G}^{(1)'}_{\Omega_{\bkg{\chi}} \sigma^*}  \, , \nonumber \\
    & \chi 
    \rightarrow \bkg{\chi} + q_\chi -
    \bkg{\sigma} ~ \overline{\G}^{(1)'}_{\Omega_{\bkg{\sigma}} \chi^*} - \bkg{\chi}~ \overline{\G}^{(1)'}_{\Omega_{\bkg{\chi}} \chi^*}  \, .
    \label{bkg.qnt.split}
\end{align}
Once applied to the tree-level vertex
functional $\G^{(0)}$,
such redefinitions generate the
linear terms in the background fields
in the r.h.s. of Eq.(\ref{bkg.1loop.feyn}).

We emphasize that the kernels $\overline{\G}^{(1)'}_{\Omega_\Phi \Phi}$ 
 depend on the fields and external sources
in a complicated way, so that Eq.(\ref{bkg.qnt.split}) is a highly non-linear redefinition w.r.t. the quantum fields.

In  the 
limit $z \rightarrow 0$
the kernels  Eq.(\ref{kernels}) reduce
to a constant:
$$
\overline{\G}^{(1)'}_{\Omega_{\bkg{\sigma}} \sigma^*} = \gamma_{\Omega_{\bkg{\sigma}} \sigma^*} \, , \quad 
\overline{\G}^{(1)'}_{\Omega_{\bkg{\chi}} \sigma^*} = 0 \, ,
\quad
\overline{\G}^{(1)'}_{\Omega_{\bkg{\sigma}} \chi^*} = 0 \, ,
\quad
\overline{\G}^{(1)'}_{\Omega_{\bkg{\chi}} \chi^*} = 
\left . \gamma_{\Omega_{\bkg{\chi}} \chi^*} \right |_{z=0} = 0 \, ,
$$
so in this limit  Eq.(\ref{bkg.qnt.split}) implies  that the 
background-quantum splitting is modified linearly, as expected by power-counting
renormalizability of the theory at $z=0$.

\subsection{Landau gauge}

At variance with the Feynman gauge, in the Landau gauge $\xi=0$ all  
the four kernels are identically zero 
since there are no interaction vertices involving the
background ghosts $\Omega$'s

As a consequence the classical background-quantum splitting
$\Phi = \bkg{\Phi} + Q_\Phi$ does not receive any radiative corrections.
This holds true 
to all orders in the loop expansion.
Thus the dependence on the background fields only originates from the 
(undeformed) background-quantum splitting.


\section{Generalized field redefinitions in the BFM}\label{sec.bgfrs}

We are now in a position to study the generalized
field redefinitions (GFRs) arising at one loop order in the presence of the background. Together 
with the renormalization of the coupling constants they allow to recursively remove
the UV divergences of the theory together with the coupling constants renormalization.

In particular the background
generalized field redefinitions
(BGFRs) can be obtained in a straightforward way by changing the 
variables from the $(A_\mu,\sigma,\chi)$-basis to the $Q_\Phi$-basis
in Eq.(\ref{bkg.1loop.feyn})
and then setting the
quantum fields to zero.

Let us start from the second term in the r.h.s. of Eq.(\ref{bkg.1loop.feyn}),
that is present both in the Feynman and Landau gauge.
According to the general results of~\cite{Gomis:1995jp} and the explicit
computations of~\cite{Binosi:2019nwz,Binosi:2020unh}, the functional
$\left . \overline{\G}^{(1)'} \right |_{\bkg{\sigma}=\bkg{\chi}=0}$ 
decomposes into the sum over a
set of integrated local gauge invariant operators ${\cal I}_j$ with gauge-independent
coefficients $c_j$\footnote{The coefficients $c_j$ run over three classes of invariants
in the classification of~\cite{Binosi:2019nwz,Binosi:2020unh}:
gauge-invariant
operators only depending on the fields; gauge-invariant operators
only depending on the external sources; gauge-invariant mixed operators
depending both on the external sources and the fields.}
 and
the functional $\overline{Y}^{(1)}$,
responsible for the generalized
field redefinitions of the quantum fields:
\begin{align}
\left . \overline{\G}^{(1)'} \right |_{\bkg{\sigma}=\bkg{\chi}=0} = \sum_j c_j {\cal I}_j 
+ \overline{Y}^{(1)} \, .
\label{one.loop.UV.div.part}
\end{align}
Both the coefficients $c$'s and the functional $\overline{Y}^{(1)}$ have been evaluated
in~\cite{Binosi:2020unh,Binosi:2019nwz} for operators up to dimension $6$.

To this approximation the ${\cal S}_0$-exact functional $\overline{Y}^{(1)}$  can be written as
\begin{align}
    \overline{Y}^{(1)} = {\cal S}_0 \int \d \Big [ 
    & \Big ( \rc{0} + \rc{1} \sigma + \rc{2} \sigma^2 + \rc{3} \chi^2 + \rc{0T} T_1 \Big ) {\cal Z}_{1} \nonumber \\
    & + \Big ( \trc{0} + \trc{1} \sigma + \trc{2} \sigma^2 
    + \trc{3} \chi^2 +
    \trc{4} \sigma \chi^2 \nonumber \\
    & \quad + \trc{0T} T_1  +  \trc{0TT} T_1^2 + \trc{1T} T_1 \sigma + 
    \trc{3T} T_1 \chi^2 
    \Big ) {\cal Z}_{2}
    \Big ] \, ,
    \label{cohom.triv.ol}
\end{align}
where we use the notation
\begin{align}
{\cal Z}_{1} &\equiv (\sigma^* \sigma  + \chi^* \chi);& 
{\cal Z}_{2} &\equiv (\sigma^* (\sigma + v) + \chi^* \chi) \, .
\label{z.combs}
\end{align}
The coefficients $\rho$'s and 
$\trc{}$'s are gauge-dependent and have been explicitly
evaluated in~\cite{Binosi:2020unh}.
In the Feynman gauge the operators arising in the functional $\overline{Y}^{(1)}$ are not gauge invariant, while in the Landau gauge they are.

In order to prove this result let us notice that 
the combination ${\cal Z}_2$ in Eq.(\ref{z.combs})
is indeed gauge invariant, since
(we can drop $b$-dependent terms by
Eq.(\ref{n.beq})):
\begin{align}
{\cal Z}_2 & = \phi \frac{\delta \G^{(0)}}{\delta \phi} +
 \phi^\dagger \frac{\delta \G^{(0)}}{\delta \phi^\dagger}
 \nonumber \\
 & =
  -2 \phi^\dagger D^2 \phi -
    \frac{2 m^2}{v^2} \Big ( \phi^\dagger \phi - \frac{v^2}{2} \Big ) \phi^\dagger \phi 
    -
    \partial^\mu T_1 ( \phi^\dagger D_\mu \phi + \mathrm{h.c.}) - T_1 (\phi^\dagger D^2 \phi + \mathrm{h.c.}) \nonumber \\
    & \quad +
    2 \bar c^* \phi^\dagger \phi + i e \phi^*  \phi \omega
    - i e (\phi^\dagger)^*  \phi^\dagger \omega
 \, ,
 \label{z2}
\end{align}
where we have introduced the notation
\begin{align}
    \phi^* = \frac{1}{\sqrt{2}}
    (\sigma^* - i \chi^{* '}) \, .
\end{align}
Then by using the values of the coefficients 
computed in~\cite{Binosi:2020unh} one obtains
\begin{align}
  \left . \overline{Y}^{(1)}  \right |_{\xi= 0 } & = 
  {\cal S}_0  \int \d \frac{M_A^2}{32 \pi^2 v^2}
  \frac{1}{\epsilon}
  \Big [  2  - 4 T_1 + 4 T_1^2 
  \nonumber \\
  & \qquad - \frac{4}{v^2} \frac{z}{1+z}
  \Big ( \phi^\dagger \phi - \frac{v^2}{2} \Big ) + 
  \frac{2}{v^4}
  \frac{z ( 3z -1)}{(1+z)^2}
  \Big ( \phi^\dagger \phi - \frac{v^2}{2} \Big )^2 
 \Big ] {\cal Z}_{2} \nonumber \\
 &  = 
 \int \d \frac{M_A^2}{32 \pi^2 v^2}
  \frac{1}{\epsilon}
  \Big [  2  - 4 T_1 + 4 T_1^2 
    - \frac{4}{v^2} \frac{z}{1+z}
  \Big ( \phi^\dagger \phi - \frac{v^2}{2} \Big ) + 
  \frac{2}{v^4}
  \frac{z ( 3z -1)}{(1+z)^2}
  \Big ( \phi^\dagger \phi - \frac{v^2}{2} \Big )^2  
 \Big ] 
 \nonumber \\
 & \qquad \qquad
 \times
 \Big [ -2 \phi^\dagger D^2 \phi -
    \frac{2 m^2}{v^2} \Big ( \phi^\dagger \phi - \frac{v^2}{2} \Big ) \phi^\dagger \phi 
    \nonumber \\ & 
    \qquad \qquad \quad
    - \partial^\mu T_1 ( \phi^\dagger D_\mu \phi + \mathrm{h.c.}) - T_1 (\phi^\dagger D^2 \phi + \mathrm{h.c.}) +
    2 \bar c^* \phi^\dagger \phi \Big ] + \dots
 \label{Y1.landau}
\end{align}
where the dots stand for additional terms of dimension $\geq 6$, $X_{1,2}$-dependent
terms (that are recovered by the replacement
in Eq.(\ref{nth.X12.eqs})) and antifield-dependent terms
that we do not need to consider.

The r.h.s. of Eq.(\ref{Y1.landau}) is gauge-invariant
by inspection, as anticipated.

We also notice that  in Landau gauge
there is a combined field renormalization
for $\sigma + v$, as a consequence
of the rigid global U(1) invariance holding true in this gauge~\cite{Sperling:2013eva}.

In the Feynman gauge instead the functional $\overline{Y}^{(1)}$  
 reads
\begin{align}
    \left . \overline{Y}^{(1)} \right |_{\xi=1} & = 
    {\cal S}_0 \int \d  \frac{M_A^2}{8 \pi^2 v^2}
    \frac{1}{\epsilon} \Big \{ \Big [ \frac{1}{1+z} 
    - \frac{2z}{(1+z)^2} \frac{\sigma}{v} + 
    \frac{z (3z -1) M_A^2}{(1+z)^3 v^2}\frac{\sigma^2}{v^2} - \frac{z}{(1+z)^2}\frac{\chi^2}{v^2}\Big ] {\cal Z}_1 \nonumber \\
    & \qquad \qquad + \frac{z}{2(1+z)} \frac{\chi^2}{v^2} {\cal Z}_2 
    \Big \} \nonumber \\
    & = \int \d \frac{M_A^2}{8 \pi^2 v^2}
    \frac{1}{\epsilon} 
    \Big \{ 
     \frac{z v}{2(1+z)} \frac{\chi^2}{v^2} \G^{(0)}_\sigma
    + \nonumber \\
    &  \qquad \qquad  + \Big [ \frac{1}{1+z} 
    - \frac{2z}{(1+z)^2} \frac{\sigma}{v} + 
    \frac{z (3z -1) M_A^2}{(1+z)^3 v^2}\frac{\sigma^2}{v^2} + \frac{z(z-1)}{2 (1+z)^2}\frac{\chi^2}{v^2}\Big ] ( \sigma \G^{(0)}_\sigma + \chi \G^{(0)}_\chi ) 
    \Big \} + \dots
    \label{y1.feyn}
\end{align}
where again the dots stand for additional terms
not contributing to the renormalization of physical operators
with dimension $\leq 6$ or $X_{1,2}$-dependent contributions.
The r.h.s of Eq.(\ref{y1.feyn}) is not gauge-invariant,
as can be directly seen.

We now collect all the factors
contributing to the classical equations of motion for $\sigma,\chi$
in Eq.(\ref{bkg.1loop.feyn}). 
There are two types of contributions:
\begin{itemize}
    \item one is associated with the deformation of the 
    background-quantum splitting at one loop order (the first term 
    between square brackets in
    the r.h.s. of Eq.(\ref{bkg.1loop.feyn}));
    \item the second is induced by the GFRs of the quantum fields
    (described by the functional $\overline{Y}^{(1)}$). It is convenient
    to parameterize $\overline{Y}^{(1)}$ as 
    \begin{align}
        \overline{Y}^{(1)} = \int \d 
        \Big ( 
        F^{(1)}_\sigma \G^{(0)}_\sigma + F^{(1)}_\chi \G^{(0)}_\chi \Big )
        + \dots \, ,
        \label{y1.eoms}
    \end{align}
    where the coefficients of the classical equations of motion $F^{(1)}_\sigma,F^{(1)}_\chi$ are gauge-dependent functionals depending
    on the fields and the external sources
    and the dots stand for
    antifield-dependent contributions that do not
    matter for the present discussion.
\end{itemize}
In order to make the connection with the usual BFM formalism,
we eventually  switch to the $(Q_\mu, q_\sigma, q_\chi)$-basis.
By looking at the coefficients
of $\G^{(0)}_\sigma,\G^{(0)}_\chi$ in Eqs.(\ref{bkg.1loop.feyn}) and
(\ref{y1.eoms})
we derive
the full renormalization of the fields, encoded in the 
following equations
\begin{align}
    & \sigma_{R} = 
   \bkg{\sigma} + q_\sigma 
      -  \bkg{\sigma} ~ \overline{\G}^{(1)'}_{\Omega_{\bkg{\sigma}} \sigma^*}- \bkg{\chi} ~
 \overline{\G}^{(1)'}_{\Omega_{\bkg{\chi}} \sigma^*} + F^{(1)}_\sigma \, , \nonumber \\
    & \chi_R = \bkg{\chi} + q_\chi 
    \rightarrow \bkg{\chi} + q_\chi -
    \bkg{\sigma} ~ \overline{\G}^{(1)'}_{\Omega_{\bkg{\sigma}} \chi^*} - \bkg{\chi}~ \overline{\G}^{(1)'}_{\Omega_{\bkg{\chi}} \chi^*} + F^{(1)}_\chi \, .
    \label{bgfrs}
\end{align}
The BGFRs are obtained from the r.h.s. of the above equation after setting
$Q_\Phi=0$ (or equivalently $A_\mu=\bkg{A}_\mu, 
\sigma=\bkg{\sigma},\chi=\bkg{\chi}$).

It turns out that such BGFRs are background-gauge invariant (although 
the kernels and the $F$-contributions in Eq.(\ref{bgfrs}) are not separately
background gauge-invariant).

At variance with the power-counting renormalizable
case, the BGFRs are non-linear.


\subsection{Background gauge invariance at $Q_\Phi=0$}

Let us now check that at zero quantum fields
one recovers a background gauge-invariant vertex
functional in agreement with the background Ward identity Eq.(\ref{bkg.wi.zero.qnt}).

In the Landau case this is obvious by inspection
since in that gauge  
$\left . \overline{Y}^{(1)} \right |_{\xi=0}$ is
separately gauge invariant
while the kernels $\G^{(1)}_{\Omega_{\bkg{\Phi} \Phi^*}}$ are
vanishing, so the whole r.h.s. of 
Eq.(\ref{one.loop.UV.div.part}) is gauge-invariant
and there are no contributions from the quantum deformation
of the background-quantum splitting.

On the other hand, in the Feynman gauge the functional 
$\left . \overline{Y}^{(1)} \right |_{\xi=1}$ is not
background gauge invariant
at $Q_\Phi=0$. Background gauge invariance
is only recovered for the sum 
(\ref{bkg.1loop.feyn}) once the contribution from the kernels
is taken into account.

In fact, as shown in Appendix~\ref{app.bkg.gfr}, 
 once one sets to zero the quantized fields
the UV divergent part of the 1-PI vertex functional 
in the Feynman gauge
reduces to
\begin{align}
\left . \overline{\G}^{(1)'}_{\xi=1} \right |_{Q_\Phi=0} = & \sum_j c_j {\cal I}_j 
- \int \d 
\frac{M_A^2}{8 \pi^2v^4} \frac{1}{\epsilon} 
\Big \{ 
\frac{z}{1+z} \Big ( \bkg{\phi}^\dagger \bkg{\phi} - \frac{v^2}{2} \Big ) 
- \frac{1}{2v^2}\frac{z (3z-1)}{(1+z)^2}
\Big ( \bkg{\phi}^\dagger \bkg{\phi} - \frac{v^2}{2} \Big )^2  \Big \} \left . {\cal Z}_2 \right |_{Q_\Phi=0}
\, .
\label{one.loop.UV.div.part.feyn}
\end{align}
Again 
by using Eq.(\ref{z2}) we see that the above
expression is gauge-invariant, as expected.

A comment is in order here. 
By comparing Eq.(\ref{one.loop.UV.div.part.feyn}) with 
Eq.(\ref{Y1.landau}) 
we see that the coefficients
of $\bkg{\phi}^\dagger \bkg{\phi} - v^2/2$
and of $(\bkg{\phi}^\dagger \bkg{\phi} - v^2/2)^2$
coincide, while the constant term
is vanishing in Feynman gauge.

The difference can be traced back
to the gauge dependence of the 
tadpole renormalization,
as discussed in Appendix~\ref{app.tadpole},
and offers an interesting example of a more
 general issue. While the functional
 $\left . \overline{\G}^{(1)'}_{\xi=1} \right |_{Q_\Phi=0}$
 is background gauge-invariant, in agreement with
 the background Ward identity Eq.(\ref{bkg.wi.zero.qnt}), 
 this does not mean that the coefficients of the
 local gauge-invariant operators in 
  $\left . \overline{\G}^{(1)'}_{\xi=1} \right |_{Q_\Phi=0}$
  are also gauge-independent. 
  It turns out that such a gauge independence only
  holds modulo the equations of motion of the theory,
  i.e. (from a cohomological point of view)
  only modulo ${\cal S}_0$-exact terms that
  are accounted for by the BGFRs.
 
\section{GFRs in the target theory}
\label{sec.gfrs.target}

The final form of the background-quantum splitting
in the target theory can be eventually read off
from Eq.(\ref{bgfrs}) by applying the mapping
in Eqs.(\ref{mapping}).
The BGFRs in the target theory are recovered by setting afterwards
$Q_\Phi=0$.

Several comments are in order.
First of all the coefficient
$F^{(1)}_\sigma$ at zero background and zero quantum fields represents the 
renormalization of the v.e.v.
Since in the Landau gauge $F^{(1)}$ is proportional to $\sigma + v$, as a consequence of the fact
that  only the invariant
${\cal Z}_2$ enters in Eq.(\ref{Y1.landau}), we conclude that no independent renormalization of the v.e.v. is present in the Landau gauge.
This is a well-known result is power-counting renormalizable theories~\cite{Sperling:2013eva}
that extend to the EFT case, being
a consequence of the rigid global U(1)
symmetry holding true in this gauge.

In the approximation 
of Eqs.(\ref{kernels}) (linear in the source $T_1$) 
the $\sigma,\chi$ redefinitions
in the presence of the backgrounds
$\bkg{\sigma},\bkg{\chi}$ read
:
\begin{align}
    \sigma_R & = \bkg{\sigma} + q_\sigma  
    - \left . \frac{M_A^2 (1 - \delta_{\xi;0})}{8 \pi^2 v^2} \frac{1}{\epsilon}
    \Big \{ \Big [ 1 - \frac{z}{1+z} \frac{\chi^2}{v^2} - \frac{g_1 v^2}{\Lambda^2}
    \Big ( \frac{1}{2} \sigma^2 + v \sigma + \frac{1}{2} \chi^2 \Big ) \Big ] \bkg{\sigma} \right . \nonumber \\
    & \qquad \qquad \qquad \qquad \quad +
    \left . \Big [ \frac{z}{1+z} - \frac{z(z-1)}{(1+z)^2}\frac{\sigma}{v}\Big ] \frac{\chi}{v} \bkg{\chi}
    \Big \} \right |_{\tiny \begin{matrix}\chi=\bkg{\chi} + q_\chi\\\sigma=\bkg{\sigma}+q_\sigma\end{matrix}} \nonumber \\
    & \qquad   + 
    \frac{M_A^2}{16 \pi^2 v} \frac{1}{\epsilon}
    \Big \{ \delta_{\xi;0} 
    + \Big [  \Big ( \frac{1-z}{1+z} - \frac{2 g_1 v^2}{\Lambda^2}  \Big ) \delta_{\xi;0} + \frac{2 \delta_{\xi;1}}{1+z}  \Big ] \frac{\sigma}{v} 
    - \Big [ \frac{4 z}{(1+z)^2} +  \frac{g_1 v^2}{\Lambda^2}
    \Big ( 3 \delta_{\xi;0} + \frac{\delta_{\xi;1}}{(1+z)^2} \Big ) \Big ] \frac{\sigma^2}{v^2} \nonumber \\
        & \qquad \qquad \qquad \qquad \quad
    + \left . \Big [ \frac{z}{(1+z)} ( \delta_{\xi;1} - \delta_{\xi;0} )
    - \frac{g_1 v^2}{\Lambda^2}\delta_{\xi;0}
    \Big ] \frac{\chi^2}{v^2} \right . \nonumber \\
    & \qquad \qquad \qquad \qquad \quad
    +  \Big [ \frac{z}{(1+z)^3}\Big ( 2  (z^2 - 1) \delta_{\xi;0} + 2 (3z-1) \delta_{\xi;1} \Big )
    - \frac{g_1 v^2}{\Lambda^2} \Big ( \delta_{\xi;0} +  \frac{\delta_{\xi;1}}{(1+z)^2} \Big ) \Big ]
     \frac{\sigma^3}{v^3}
    \nonumber \\
     & \qquad \qquad \qquad \qquad \quad   
    + \left . \Big [ - \frac{2z}{1+z} \Big ( \frac{1-z}{1+z}\delta_{\xi;0}
    + \delta_{\xi;1}\Big ) 
    - g_1 \Big ( \delta_{\xi;0} + \frac{\delta_{\xi;1}}{(1+z)^2}
    \Big ) 
    \Big ] \frac{\chi^2 \sigma}{v^3} \right . \nonumber \\
    & \qquad \qquad \qquad \qquad \quad  
    \left . 
    + \frac{z (-1)^{\delta_{\xi;1}} }{(1+z)^2}
    \Big ( 3z + (-1)^{\delta_{\xi;0}}  \Big ) \frac{\chi^2 \sigma^2}{v^4}
    \Big ]
    \Big \} \right |_{\tiny \begin{matrix}\chi=\bkg{\chi} + q_\chi\\\sigma=\bkg{\sigma}+q_\sigma\end{matrix}} + \dots \, ,
    \nonumber \\
    \chi_R & = \bkg{\chi} + q_\chi - 
     \left . \frac{M_A^2 (1 - \delta_{\xi;0})}{8 \pi^2 v^2} \frac{1}{\epsilon}
    \Big \{ \Big [ \frac{z}{1+z} -
\frac{z(z-1)}{(1+z)^2}\frac{\sigma}{v} \Big ] \frac{\chi}{v}  \bkg{\sigma} 
     \right .
     \nonumber \\ 
     & \qquad \qquad \qquad \qquad  +
    \left . \Big [\frac{1}{1+z} - \frac{2z}{(1+z)^2}\frac{\sigma}{v} +
    \frac{z(3z-1)}{(1+z)^3}\frac{\sigma^2}{v^2}
    + \frac{z^2}{(1+z)^2}\frac{\chi^2}{v^2} \right . \nonumber \\
    & \qquad \qquad \qquad \qquad \quad
    \left . - \frac{1}{(1+z)^2} \frac{g_1 v^2}{\Lambda^2}
    \Big ( \frac{1}{2} \sigma^2 + v \sigma + \frac{1}{2}\chi^2 \Big )
    \Big ]
    \bkg{\chi}
    \Big \} \right |_{\tiny \begin{matrix}\chi=\bkg{\chi} + q_\chi\\\sigma=\bkg{\sigma}+q_\sigma\end{matrix}} \nonumber \\
    & 
    \qquad \qquad  + 
    \frac{M_A^2}{16 \pi^2 v} \frac{1}{\epsilon}
    \Big \{ 
    \delta_{\xi;0} + \frac{2}{1+z}
    \delta_{\xi;1}
    - 
    \Big [\frac{2 z (1+z) \delta_{\xi;0} + 4 z \delta_{\xi;1} }{(1+z)^2} 
    + \frac{2 g_1 v^2}{\Lambda^2}\Big ( \delta_{\xi;0}
    + \frac{\delta_{\xi;1}}{(1+z)^2} \Big )
    \Big ]
     \frac{\sigma}{v}
     \nonumber \\
     &
     \qquad \qquad \qquad \quad
     +  \Big [ \frac{2 z}{(1+z)^3}
     \Big ( (z^2-1) \delta_{\xi;0} + 
     (3z - 1)\delta_{\xi;1} \Big ) 
     - \frac{g_1 v^2}{\Lambda^2} 
     \Big (
     \delta_{\xi;0} + 
     \frac{\delta_{\xi;1}}{(1+z)^2}
     \Big )
     \Big ]
     \frac{\sigma^2}{v^2}
    \nonumber \\
    &
    \qquad \qquad \qquad \quad
    \left . - \Big [ \frac{z}{(1+z)^2}\Big ( 
    (1+z) \delta_{\xi;0} + (1-z)\delta_{\xi;1}
    \Big ) + \frac{g_1 v^2}{\Lambda^2}
    \Big ( 
    \delta_{\xi;0} +  \frac{\delta_{\xi;1}}{(1+z)^2} \Big )
    \Big ] \frac{\chi^2}{v^2} \right . \nonumber \\
        &
    \qquad \qquad \qquad \quad
    \left . 
    + \frac{(-1)^{\delta_{\xi;1}} z}{(1+z)^2} 
    \Big [ 3z + (-1)^{\delta_{\xi;0}} \Big ]
     \frac{\sigma \chi^2}{v^3}
    \Big \} \frac{\chi}{v} 
     \right |_{\tiny \begin{matrix}\chi=\bkg{\chi} + q_\chi\\\sigma=\bkg{\sigma}+q_\sigma\end{matrix}}
    + \dots
    \label{target.gfrs}
\end{align}
The dots stand for terms cubic in $\sigma$ 
or of dimension $\geq 4$ that do not contribute to the renormalization of dim.6 operators~\cite{Binosi:2020unh}.

The BGFRs are obtained
by setting $q_\chi=q_\sigma=0$
in Eq.(\ref{target.gfrs}). 
As already noticed, in this limit
the vertex functional becomes background gauge-invariant
w.r.t. the variation of the background fields only.
Accordingly the BGFRs are generated by a multiplicative
redefinition of the background fields by 
a gauge-invariant polynomial that can be immediately
read off from Eq.(\ref{param}) :
\begin{align}
    \begin{pmatrix}
        \bkg{\sigma}_R  \\
        \bkg{\chi}_R 
    \end{pmatrix}
= \Big [
a_0 + a_1 T_1   
    + a_2 \Big ( \bkg{\phi}^\dagger \bkg{\phi} - \frac{v^2}{2} \Big ) +
    a_3 \Big (\bkg{\phi}^\dagger \bkg{\phi} - \frac{v^2}{2} \Big )^2 + \dots
\Big ]    
\begin{pmatrix}
        \bkg{\sigma} + v \\
        \bkg{\chi}
    \end{pmatrix}
\end{align}
with the coefficients $a$'s given by Eq.(\ref{bgfr.param}).
Notice that these coefficients are in general gauge-dependent.

\section{Conclusions}\label{sec.concls}

In the present paper we have investigated 
the renormalization 
of the quantum and background fields in a spontaneously
broken gauge effective field theory. We have shown that in this
class of models, where power-counting renormalizability is lost,
both the background and the quantum field renormalize in a 
non-linear way.

One must take into account the 
contributions from the radiative
deformation of the classical
background-quantum splitting as well as
the effect of the non-linear GFRs
of the quantum fields.

At zero quantum fields $Q_\Phi=0$
one recovers background gauge invariance
of the vertex functional w.r.t
the transformation of the background fields.
This property is reflected on the
background gauge-invariance of the
background-dependent counter-terms.

However, despite such background gauge invariance
at zero quantum fields, the coefficients of the background invariants that are 
proportional to the equations of motion are in general gauge-dependent.

Consequently the correct renormalization of gauge-invariant operators requires to take into account the effect of the BGFRs already at one loop order.

For higher order computations, the much more complicated 
background and quantum
generalized field redefinitions must be carried out in order to achieve the symmetric
subtraction
of the theory under consideration.

The tools and results  in the present paper pave the way to further applications to non-Abelian effective
gauge theories and in particular to the SMEFT.

\section*{Acknowledgments}

Useful discussions with D.Anselmi and D.Binosi are gratefully acknowledged.

\appendix


\section{Symmetries of the theory}\label{app.f.ids}

Several functional identities hold for $\G$:
\begin{itemize}
    \item the  Slavnov-Taylor (ST) identity associated with the gauge BRST symmetry
    
    %
    %
    %
    The ST identity for the  vertex functional $\G$ 
    generated by the gauge BRST differential $s$ reads
    \begin{align}
	& {\cal S}(\G)  = \int \mathrm{d}^4x \, \Big [ 
	\partial_\mu \omega \frac{\delta \G}{\delta A_\mu} + \frac{\delta \G}{\delta \sigma^*} \frac{\delta \G}{\delta \sigma}  + \frac{\delta \G}{\delta \chi^*} \frac{\delta \G}{\delta \chi} 
	+ b \frac{\delta \G}{\delta \bar \omega} 
	+ \Omega_\mu \frac{\delta \G}{\delta \bkg{A}_\mu} +
	\Omega_{\bkg{\sigma}} \frac{\delta \G}{\delta \bkg{\sigma}} + \Omega_{\bkg{\chi}} \frac{\delta \G}{\delta \bkg{\chi}}
	\Big ] = 0 \, .
	\label{sti} 
	\end{align}
\item the constraint ST identity
%
 
The ST identity associated with the BRST differential $\s$ is
\begin{align}
    {\cal S}_C (\G) = \int \d \Big [ vc \frac{\delta \G}{\delta X_1} + \frac{\delta \G}{\delta \bar c^*} \frac{\delta \G}{\delta \bar c} \Big ] = 
    \int \d \Big [vc \frac{\delta \G}{\delta X_1} -
    (\square + m^2) c \frac{\delta \G} {\delta \bar c^*}  \Big ] = 0 \, ,
    \label{sti.c}
\end{align}
where in the second term of the above equation we have used the fact that the fields $c,\bar c$ are free:
\begin{align}
    \frac{\delta \G}{\delta \bar c} = - (\square + m^2) c \, ; \qquad \qquad 
    \frac{\delta \G}{\delta c} =  (\square + m^2) c \, .
    \label{c.eoms}
\end{align}
\item the $X_{1,2}$-equations

Since $c$ is a free field, the constraint
ST identity Eq.~(\ref{sti.c}) reduces to the $X_1$-equation of motion
\begin{align}
{\cal B}_{X_1}(\G) \equiv
    \frac{\delta \G}{\delta X_1} -
    \frac{1}{v} (\square + m^2)
    \frac{\delta \G}{\delta \bar c^*} = 0\, .
    \label{X1.eq}
\end{align}
The $X_2$-equation is in turn given by
\begin{align}
{\cal B}_{X_2}(\G) \equiv &
	\frac{\delta \G}{\delta X_2} -  \frac{1}{v} (\square + m^2) \frac{\delta \G}{\delta \bar c^*} - \frac{g_1 v}{ \Lambda^2} \frac{\delta \G}{\delta T_1} 
	- \frac{g_2 v}{\Lambda^2} \frac{\delta \G}{\delta U} 
	- \frac{g_3 v^3}{2 \Lambda^2} \frac{\delta \G}{\delta R}
	\nonumber \\
	& = - (\square + m^2)X_1
	- \Big [ (1+z) \square + M^2 \Big ] X_2 - v \bar c^* .
	\label{X2.eq}
\end{align}
Both Eqs.(\ref{X1.eq}) and (\ref{X2.eq}) are unaltered
by the presence of the background fields.
At order $n$, $n \geq 1$ in the loop expansion the $X_{1,2}$-equations 
read
\begin{subequations}
\begin{align}
    \frac{\delta \G^{(n)}}{\delta X_1} &=
    \frac{1}{v} (\square + m^2)
    \frac{\delta \G^{(n)}}{\delta \bar c^*}, \\
    \frac{\delta \G^{(n)}}{\delta X_2} &=  \frac{1}{v} (\square + m^2) \frac{\delta \G^{(n)}}{\delta \bar c^*} + \frac{g_1 v}{ \Lambda^2} \frac{\delta \G^{(n)}}{\delta T_1} 
	+ \frac{g_2 v}{ \Lambda^2} \frac{\delta \G^{(n)}}{\delta U} 
	+ \frac{g_3 v^3}{2 \Lambda^2} \frac{\delta \G^{(n)}}{\delta R}.
\end{align}
	\label{nth.X12.eqs}
\end{subequations}

By using the chain rule for functional differentiation we see that by Eqs.~(\ref{nth.X12.eqs}) $\G^{(n)}$ can only depend on the combinations:
\begin{subequations}
\begin{align}
	\tbarc &= \bar{c}^* + \frac{1}{v}(\square + m^2) (X_1 + X_2);& \tT &= T_1 + \frac{g_1 v}{\Lambda^2}X_2, \nonumber \\
	\tU &= U + \frac{g_2 v}{\Lambda^2}X_2;& \tR &= R + \frac{g_3 v^3}{2\Lambda^2}X_2.
    \label{X12.sols}
\end{align}
\end{subequations}
Notice that 
the combinations in the r.h.s.
of Eq.(\ref{X12.sols}) are background gauge invariant.
\item the $b$-equation
\begin{align}
    \frac{\delta \G}{\delta b} = \xi b - 
    \partial^\mu (A_\mu - \bkg{A}_\mu ) - \xi e [ (\bkg{\sigma} + v)\chi -
    \bkg{\chi} (\sigma + v) ] \, .
    \label{b.eq}
\end{align}
By projecting 
Eq.(\ref{b.eq})
at order $n\geq 1$  one sees that, as usual in linear gauges, the
$b$-dependence is confined at tree level:
\begin{align}
    \frac{\delta \G^{(n)}}{\delta b} = 0 \, , \qquad n \geq 1 \, .
    \label{n.beq}
\end{align}
Hence in studying  higher order 1-PI Green's functions one can safely set $b=0$.
\item the antighost equation
\begin{align}
 \frac{\delta \G}{\delta \bar \omega} =
 \square \omega + \xi e v \frac{\delta \G}{\delta \chi^*}
 - \partial^\mu \Omega_\mu + \xi e \Omega_{\bkg{\sigma}} \chi
 - \xi e \Omega_{\bkg{\chi}} (\sigma + v) \, .
\end{align}
At order $n \geq 1$
the above equation reads
\begin{eqnarray}
     \frac{\delta \G^{(n)}}{\delta \bar \omega} = \xi e v \frac{\delta \G^{(n)}}{\delta \chi^*} \, .
     \label{n.ag}
\end{eqnarray}
Eq.(\ref{n.ag}) entails
that at order $n \geq 1$
the dependence on $\bar\omega$
only happens via the combination
\begin{align}
    {\chi^{*}}^\prime = \chi^* + \xi ev \bar \omega \, .
    \label{redef.chistar}
\end{align}
\item the background Ward identity
\begin{align}
    {\cal W} (\Gamma ) = &
      - \partial^\mu \frac{\delta \G}{\delta A^\mu}
    - e \chi \frac{\delta \G}{\delta \sigma} 
    + e (\sigma + v) \frac{\delta \G}{\delta \chi} 
      - \partial^\mu \frac{\delta \G}{\delta \bkg{A}^\mu}
    - e \bkg{\chi} \frac{\delta \G}{\delta \bkg{\sigma}} 
    + e (\bkg{\sigma} + v) \frac{\delta \G}{\delta \bkg{\chi}} \nonumber \\
    & 
     - e \chi^* \frac{\delta \G}{\delta \bkg{\sigma^*}} 
    + e \sigma^* \frac{\delta \G}{\delta \chi^*} = 0 \, .
    \label{bkg.wi}
\end{align}
\end{itemize}

\section{Parameterization
of Background Generalized Field Redefinitions}\label{app.bkg.gfr}

We parameterize the terms proportional to the classical equations of motion
for $\sigma,\chi$ in Eq.(\ref{bkg.1loop.feyn})
at $Q_\Phi=0$ (i.e. $\bkg{A}_\mu=A_\mu,\bkg{\sigma}=\sigma,
\bkg{\chi}=\chi$)
as follows
\begin{align}
    & \int \d \Big \{ a_0 + a_1 T_1   
    + a_2 \Big ( \bkg{\phi}^\dagger \bkg{\phi} - \frac{v^2}{2} \Big ) +
    a_3 \Big (\bkg{\phi}^\dagger \bkg{\phi} - \frac{v^2}{2} \Big )^2 + \dots
     \Big \} \left .
    {\cal Z}_2  \right |_{Q_\Phi=0} \,  \nonumber \\
    & = \int \d \Big [ \Big ( r + r_{T_1} T_1 + r_\sigma \bkg{\sigma} + r_{\sigma^2} \bkg{\sigma}^2 
    + r_{\chi^2} \bkg{\chi}^2 
    +\dots \Big ) \left . \G^{(0)}_\sigma \right |_{Q_\Phi=0} 
    \nonumber \\
    & \qquad \quad ~ + 
    \Big ( r_\chi \bkg{\chi} 
    + r_{\chi T_1} \bkg{\chi} T_1
    + r_{\chi\sigma} \bkg{\chi} \bkg{\sigma} 
    + r_{\chi\sigma^2} \bkg{\chi} \bkg{\sigma}^2  
    + r_{\chi^3} \bkg{\chi}^3 
    +\dots \Big ) \left . \G^{(0)}_\chi \right |_{Q_\Phi=0}
    \Big ] \, ,
    \label{param}
\end{align}
where the dots denote terms at least quadratic in $T_1$,
cubic in $\sigma$ or of dimension $\geq 4$
that can be neglcted in the approximation of the GFRs used in~\cite{Binosi:2020unh}.

The coefficients $r$'s are  known since 
they are linear combinations of the $\tilde{\rho}$'s in
Eq.(\ref{cohom.triv.ol})
and the $\gamma$'s in the kernel expansion
Eq.(\ref{exp.kernel}), namely 
\begin{eqnarray}
    & r = v \tilde{\rho}_0 \, , \quad r_\sigma = \rho_0 + \tilde{\rho}_0 + v \tilde{\rho}_1  - \gamma_{\Omega_{\bkg{\sigma}} \sigma^*} \, , \nonumber \\
    &
    r_{T_1} = v \tilde{\rho}_0 \, , \quad
    r_\sigma^2 = 
    \rho_1 + \tilde{\rho}_1 + v \tilde{\rho}_2
    - \gamma_{\Omega_{\bkg{\sigma}} \sigma^* \sigma}
    - \gamma_{\Omega_{\bkg{\sigma}} \sigma^* \bkg{\sigma}} \, , 
    \quad
    r_{\chi^2} = v \tilde{\rho}_3 
    - \gamma_{\Omega_{\bkg{\chi}} \sigma^* \chi}
    - \gamma_{\Omega_{\bkg{\chi}} \sigma^* \bkg{\chi}} \, ,
    \nonumber \\
    & r_\chi = \rho_0 + \tilde{\rho}_0 
    - \gamma_{\Omega_{\bkg{\chi}} \chi^*} \, , 
    \qquad
    r_{\chi T_1} = \rho_{0T} + \tilde{\rho}_{0T} 
    - \gamma_{\Omega_{\bkg{\chi}} \chi^* T_1}\, ,
   \nonumber \\
   &
    r_{\chi\sigma} =  \rho_1 + \tilde{\rho}_1
      -  \gamma_{\Omega_{\bkg{\sigma}} \chi^* \chi}
      -  \gamma_{\Omega_{\bkg{\sigma}} \chi^* \bkg{\chi}}
      -  \gamma_{\Omega_{\bkg{\chi}} \chi^* \sigma}
      -  \gamma_{\Omega_{\bkg{\chi}} \chi^* \bkg{\sigma}}\, , 
      \nonumber \\
      & 
    r_{\chi\sigma^2} =  \rho_2 + \tilde{\rho}_2 
    - \frac{1}{2} \gamma_{\Omega_{\bkg{\chi}} \chi^* \sigma \sigma } 
    -\frac{1}{2} \gamma_{\Omega_{\bkg{\chi}} \chi^* \bkg{\sigma} \bkg{\sigma} } 
    - \gamma_{\Omega_{\bkg{\chi}} \chi^* \bkg{\sigma} \sigma } 
    -
    \gamma_{\Omega_{\bkg{\sigma}} \chi^* \sigma \chi} -
    \gamma_{\Omega_{\bkg{\sigma}} \chi^* \bkg{\sigma} \chi} -
    \gamma_{\Omega_{\bkg{\sigma}} \chi^* \sigma \bkg{\chi}} -
    \gamma_{\Omega_{\bkg{\sigma}} \chi^* \bkg{\sigma} \bkg{\chi}} 
    \, , \nonumber \\
    &
     r_{\chi^3} =
    \rho_3 + \tilde{\rho}_3 
    - \frac{1}{2} \gamma_{\Omega_{\bkg{\chi}} \chi^* \chi^2} 
    - \frac{1}{2} \gamma_{\Omega_{\bkg{\chi}} \chi^* \bkg{\chi}^2}
    - \gamma_{\Omega_{\bkg{\chi}} \chi^* \bkg{\chi} \chi}
\, .
\end{eqnarray}
The coefficients $a$'s can be expressed
as linear combinations of the $r$'s.
The linear system is over-constrained, so
we obtain some consistency conditions
that have to be fulfilled.

We find
\begin{subequations}
\begin{equation}
a_0 = r_\chi \, ,
\qquad 
a_1 = r_{\chi T_1} \, , \qquad
v a_2 = r_{\chi\sigma} \, , 
\qquad
\frac{a_2}{2} + v^2 a_3 = r_{\chi\sigma^2} \, ,
\label{sols}
\end{equation}
\begin{equation}
r = v a_0 \, , \quad
r_\sigma = v^2 a_2 + a_0 \, , \quad
r_{\sigma^2} = \frac{3}{2} v a_2 + 
v^3 a_3 \, , \quad 
r_{\chi^2} = \frac{v}{2} a_2 \, ,
\quad 
r_{\chi^3}=\frac{a_2}{2} \, ,
\quad
r_{T_1} = v a_1 \, .
\label{cc}
\end{equation}
\end{subequations}
Eqs.(\ref{sols}) fix the coefficients $a$'s,
while Eqs.(\ref{cc}) are the consistency
conditions that must be fulfilled.
One finds for the $a$'s:
\begin{align}
    & a_0 = \frac{M_A^2 (1- \delta_{\xi;1})}{16 \pi^2 v^2}\frac{1}{\epsilon} \, , \qquad
    a_1 = -\frac{M_A^2 (1- \delta_{\xi;1})}{8 \pi^2 v^2}\frac{1}{\epsilon} \, , \nonumber \\
    & a_2 = - \frac{1}{8 \pi^2 v^4}\frac{z M_A^2}{1+z} \frac{1}{\epsilon} \, , \qquad
    a_3 = \frac{1}{16 \pi^2 v^6} \frac{M_A^2 z (3z -1)}{(1+z)^2} \frac{1}{\epsilon} \, .
    \label{bgfr.param}
\end{align}
It is then easy to check that they obey
Eqs.(\ref{cc}).

\section{Tadpole renormalization}
\label{app.tadpole}

The coefficient $c_0$ 
in Eq.(\ref{param}) is related to the renormalization of the
tadpole $\G^{(1)}_{\sigma(0)}$. 

We begin by studying the
background taadpole.
By taking a derivative of 
the ST identity Eq.(\ref{sti}) w.r.t. $\Omega_{\bkg{\sigma}}$ 
and then setting all the fields and external sources to zero we obtain
\begin{align}
    \G^{(1)}_{\bkg{\sigma}(0)} = 0 \, ,
\end{align}
i.e. the background tadpole vanishes (in the $(A_\mu,\sigma,\chi)$-basis)
in any gauge as a consequence of the ST identity.

The UV-divergent part of the $\sigma$-tadpole $\overline{\G}^{(1)}_{\sigma(0)}$
can be read off from Eq.(\ref{one.loop.UV.div.part}):
\begin{align}
\left . \overline{\G}^{(1)'} \right |_{\bkg{\sigma}=\bkg{\chi}=0} = \sum_j c_j {\cal I}_j 
+ \overline{Y}^{(1)} \supset 
\lambda_1 \int \d \, \Big ( \phi^\dagger \phi - \frac{v^2}{2} \Big ) +
c_0 \int \d \, \Big [ (\sigma + v) \G^{(0)}_\sigma + \chi \G^{(0)}_\chi \Big ]
\, .
\label{one.loop.UV.div.part.1}
\end{align}
By taking a derivative of Eq.(\ref{one.loop.UV.div.part.1})  w.r.t $\sigma$ 
and then setting fields and external sources to zero we obtain
\begin{align}
  \overline{\G}^{(1)}_{\sigma(0)} = v \lambda_1 - m^2 v c_0 \, .
  \label{tadpole}
\end{align}
The coefficient $\lambda_1$ reads
\begin{align}
\lambda_1 = \frac{1}{16 \pi^2 v^2}
\frac{1}{(1+z)^3} 
\Big \{
(1+z) [ M^2 + M_A^2(1+z)^2] m^2  + 2 [ M^4 + 3 M_A^4 (1+z)^3]
\Big \} 
\frac{1}{\epsilon} \, ,
\end{align}
where $M_A=e v$ is the mass of the vector meson $A_\mu$.

In Feynman gauge $c_0$ vanishes so that 
\begin{align}
 \lambda_1 = \left . \frac{1}{v} \overline{\G}^{(1)}_{\sigma(0)} \right |_{\xi=1} \, .
\end{align}
Eq.(\ref{tadpole}) then implies in the Landau gauge:
\begin{align}
    \left . c_0
    \right |_{\xi=0} = \frac{1}{m^2 v} \Big (  \left .  \overline{\G}^{(1)}_{\sigma(0)} \right |_{\xi=1} - 
    \left .  \overline{\G}^{(1)}_{\sigma(0)} \right |_{\xi=0}
    \Big ) \, .
    \label{eq.c0}
\end{align}
This is a consistency relation satisfied by the  coefficient $c_0$ in Eq.(\ref{eq.c0}) that can be easily
verified by explicit computation.


%

\end{document}